\documentclass[usenatbib]{mnras}
\usepackage{amsmath}
\usepackage{multirow}  
\usepackage{graphicx}  
\usepackage{subcaption}
\usepackage{float}
\usepackage{bm}
\usepackage{comment}

\usepackage{newtxtext,newtxmath}

\usepackage[T1]{fontenc}
\usepackage{booktabs}
\usepackage{arydshln}  
\DeclareRobustCommand{\VAN}[3]{#2}
\let\VANthebibliography\thebibliography
\def\thebibliography{\DeclareRobustCommand{\VAN}[3]{##3}\VANthebibliography}

\usepackage{graphicx}	 
\usepackage{amsmath}	 
\usepackage{xcolor}

\title[Two-component gas]{A two-component profile of baryonic feedback: hydrostatic and diffuse gas}

\author[Shavelle et al.]{
Kaitlyn Shavelle,$^{1}$\thanks{E-mail: ks8924@princeton.edu}
Jared Siegel,$^{1}$
Leah Bigwood,$^{2,3}$
Alexandra Amon,$^{1}$
Romain Teyssier$^{1}$
\\
$^{1}$Department of Astrophysical Sciences, Princeton University, Princeton, NJ 08544, USA\\
$^{2}$Institute of Astronomy, University of Cambridge, Madingley Road, Cambridge, CB3 0HA, UK\\
$^{3}$Kavli Institute for Cosmology (KICC), University of Cambridge, Madingley Road, Cambridge CB3 0HA, UK\\
}

\date{Accepted XXX. Received YYY; in original form ZZZ}

\pubyear{\the\year{}}

\begin{document}
\label{firstpage}
\pagerange{\pageref{firstpage}--\pageref{lastpage}}
\maketitle

\begin{abstract}
Baryonic feedback processes redistribute gas within and beyond dark matter halos, however, the resulting gas distribution remains highly uncertain, limiting predictions of the matter distribution for precision cosmology. 
We introduce a physically motivated approach to modelling the gas distribution that decomposes it into an approximately hydrostatic inner component and a diffuse outer component.
This approach takes advantage of the growing observational landscape: X-ray measurements primarily probe the dense inner halo, while the kinetic Sunyaev–Zel’dovich (kSZ) effect is sensitive to diffuse ionized gas extending to larger radii. 
We implement this two-component profile within the baryonification framework, though we note that it is applicable to any model of the large-scale gas distribution. 
We jointly fit X-ray gas-fraction measurements from the HSC–XXL and eROSITA samples with stacked kSZ effect measurements of BOSS CMASS galaxies from the Atacama Cosmology Telescope, demonstrating that the proposed profile retains sufficient flexibility. 
This proof-of-concept shows that a decomposition between hydrostatic and diffuse gas provides a simple, physically interpretable framework for connecting complementary observations of the gas distribution, and a foundation for future models of baryonic feedback.
\end{abstract}


\section{Introduction}
Baryonic feedback processes, including supernovae, stellar winds, and particularly, active galactic nuclei (AGN) influence the matter distribution within and beyond dark matter halos through gas heating and ejection \citep[e.g.,][]{King_2015, Fabian_2012, Hlavacek-Larrondo_2022, Harrison_2024}. %By altering the matter distribution, 
These processes suppress the matter power spectrum on small, non-linear scales by tens of percent relative to dark matter-only predictions \citep[e.g.][]{Semboloni_2011,vanDaalen2011}.
Cosmological hydrodynamical simulations reproduce many observed properties of galaxies and clusters, however, different feedback implementations make widely varying predictions for the impact of baryons on large scales \citep{ Henden_2018, McCarthy_2016, Dav__2019, Schaye_2023,  Pakmor_2023}. 
As such, baryonic feedback is one of the dominant astrophysical uncertainties for precision cosmology \citep{Chisari_2019, DES_Y3, Robertson_2025, DES_Y6}.

The impact of baryonic feedback on cosmological observables is commonly modelled with a range of complementary methods \citep{Chisari_2019}, for example: %fitting functions calibrated to the matter 
simulation-based fitting functions and emulators \citep{van_Daalen_2019, Salcido_2023,Schaller2025}, halo-model prescriptions \citep{Semboloni_2011,Mead_2021,Pandey2025godmax}, baryonification methods \citep{Schneider_2015,Angulo2021, Arico2021,Schneider_2019, schneider_2025}, the resummation model \citep{van_Daalen_2025}, and phenomenological approaches \citep{Huang2021, Amon_2022, preston_2023}. These approaches have proven successful at capturing baryonic effects while remaining fast enough for cosmological inference. However, uninformed flexible models can substantially degrade the constraining power of weak lensing measurements \citep{bigwood_2024}. Incorporating additional observational constraints on the gas distribution is therefore critical to unlocking small-scale cosmological information and understanding the physics of baryonic feedback \citep[e.g.][]{Schneider_2022, BigwoodBB_2026}.

In this work, we introduce a new approach for describing the gas distribution around halos: a physically motivated two-component model consisting of a hydrostatic inner component and a flexible diffuse outer component. 
This two component approach is applicable to any method of predicting the effects of feedback on large-scale structure.
Here, we implement this idea as an extension of the baryonification framework, a procedure for modifying dark matter only profiles to mimic baryonic effects \citep{Schneider_2015}.
As a demonstration, we present a new gas density profile that can be fed into the baryonification process.  We leave a full implementation, including predictions of the matter power spectrum, to later work. 
This framework provides a simple, physically interpretable description of halo gas that is naturally suited to incorporating future constraints on the circumgalactic and intracluster mediums.

The state of the gas embedded in dark matter halos varies substantially with radius. The intracluster medium within approximately $R_{500}$ is dense, hot, and close to hydrostatic equilibrium \citep{Borgani_2011}. At larger radii (beyond several times $R_{500}$), the gas becomes increasingly diffuse and deviates from virial equilibrium \citep{Borgani_2011, Fedeli_2014}. AGN feedback is believed to play a central role in the shaping of this radial gas distribution by heating and redistributing gas from the centre of the halo into the surrounding circumgalactic and intracluster medium  
\citep{Sijacki2006,McNamara2007,Conroy2008,Booth_2009,McCarthy2011,Fabian_2012,Gitti2012,Gaspari_2020,Eckert_2021}.

\begin{figure*}
    \centering
    \includegraphics[width=2.0\columnwidth]{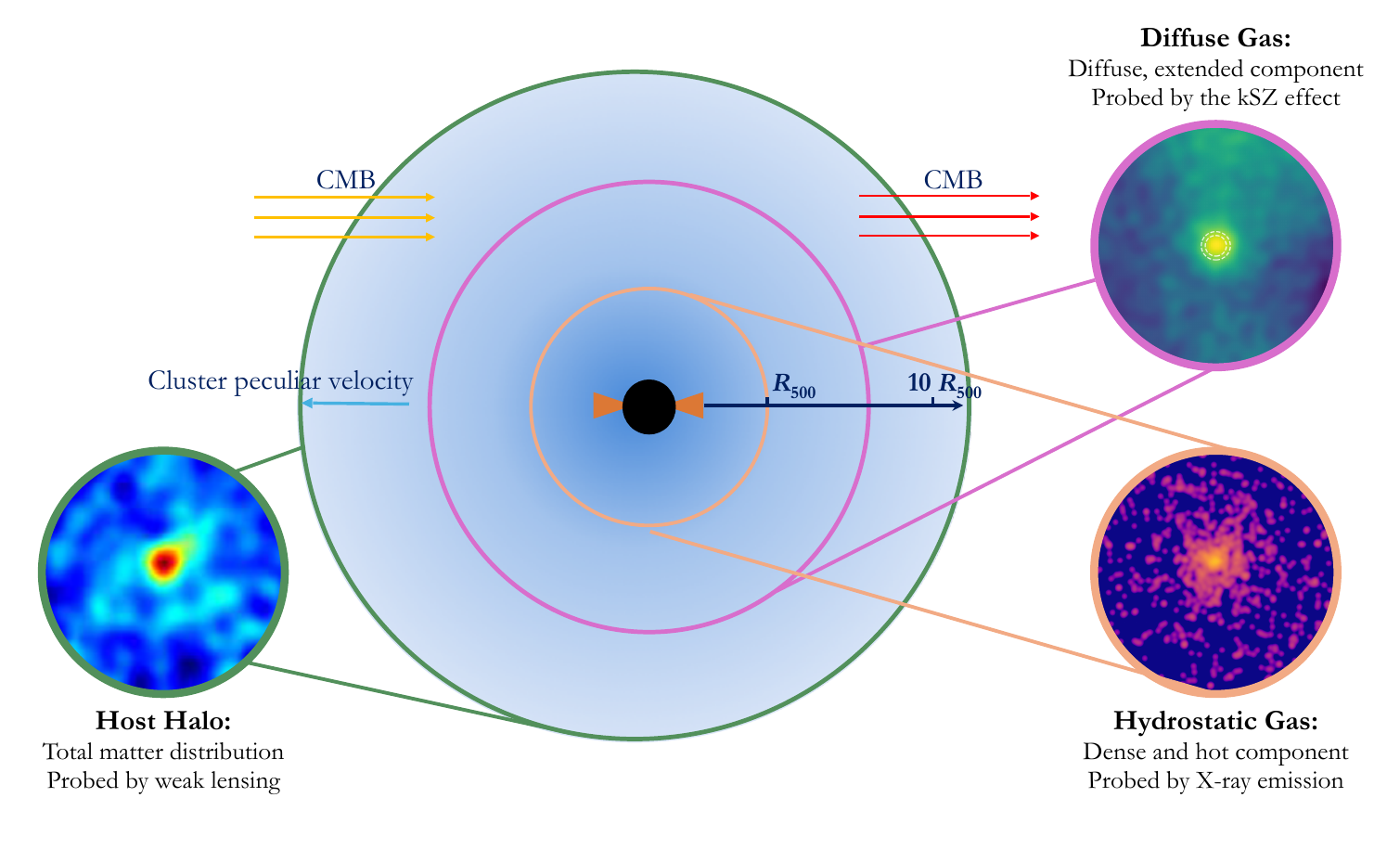}
   \caption{Schematic of the motivation for the two-component gas profile. The inner region hosts dense, hot gas near hydrostatic equilibrium, while the gas is diffuse and extended at larger radii. These two components are probed by three complementary observables: the kSZ effect traces the diffuse gas in the outer halo, X-ray gas measurements probe the denser hot gas in the inner region, and weak lensing constrains the total mass of the host halo. The X-ray, kSZ, and lensing insets are reproduced from \citet{Bublul_2024}, \citet{Qu2026}, and \citet{Umetsu_2020}, respectively.
   }
    \label{fig:schematic_figure}
\end{figure*}

Constraining the gas distribution across a wide range of radii, mass, and redshift requires a multi-probe approach \citep[e.g.][]{Battaglia_2017}.
X-ray emission depends on the squared gas density (at fixed metallicity and temperature), making it primarily sensitive to dense, hot gas in the centres of massive halos \citep[e.g.,][]{Seppi2022,Marini2024,Clerc2024}.
The kSZ effect, sourced  by the inverse Compton scattering of cosmic microwave background (CMB) photons by free electrons in groups and clusters, depends linearly on density \citep{Sunyaev_1980}.  
Relative to X-ray observations, the kSZ effect is therefore more sensitive to the diffuse ionized gas at large halo radii (beyond $R_{500}$) of lower mass halos \citep{ho_2009, Qu2026, hadzhiyska_2025}. 
The thermal SZ effect is sensitive to the electron pressure \citep[e.g.][]{Hand_2011, Siegel_2026},
while fast radio bursts offer a new probe of the electron column density through dispersion measures \citep[e.g.][]{Reischke_2026, connor_2024,  scharma_2024}. Galaxy-galaxy lensing (GGL) probes the total matter distribution of the halo \citep[e.g.][]{Mandelbaum_2013, siegel2025_flamingo}.  
Together, these observables provide highly complementary constraints on the gas distribution and its thermodynamic state (see Figure \ref{fig:schematic_figure}). 
This complementarity allows us to constrain the two-component approach proposed here.

In Section~\ref{sec:model}, we introduce the proposed two-component profile, which we refer to as \textsc{HyDif}.
The model is compared against a number of cosmological hydrodynamical simulations in Section~\ref{sec:sims} and demonstrated by jointly fitting kSZ effect and X-ray gas mass fraction measurements in Section~\ref{sec:data}.  We discuss future applications in Section~\ref{sec:discussion} and conclude in Section~\ref{sec:conclude}.

\section{Modelling the Gas distribution}\label{sec:model}
We propose a two-component profile for describing the gas distribution around halos;
we implement this profile within the baryonification framework.
In Section~\ref{sec:bfc_review}, we briefly review the baryonification framework and 
the two-component extension is described in Section~\ref{sec:two_component}.

\begin{table*}
\centering
\caption{Free parameters in the \textsc{HyDif} model. For each parameter, we list its description and its corresponding flat uniform prior and
fiducial value. All three parameters shown are varied in the fitting process:
$\log_{10}M_{\rm c}$ and $\mu$ control the inner
(hydrostatic) component through the component fraction, while
$\theta_{\rm d}$ sets the maximum radius of gas ejection of the outer
(diffuse) component. The remaining shape parameters of the model are held fixed at their fiducial values and are not listed here.}
\label{tab:bc_update}
\setlength\dashlinedash{3pt}
\setlength\dashlinegap{2pt}
\resizebox{\textwidth}{!}{%
\begin{tabular}{l p{10cm} c c }
\toprule
\textbf{Parameter} & \textbf{Description} & \textbf{Prior} & \textbf{Fiducial value} \\
\midrule
$\log_{10} M_{\rm c}$ & Characteristic mass scale above which the hydrostatic component becomes dominant; defines the component fraction: Equation~\eqref{fcomp_equation}. & U[11,\,15] & 13.32  \\
$\mu$ & Slope of the hydrostatic component fraction power-law. & U[0.0,\,2.0] & 0.5  \\
$\theta_{\mathrm{d}}$ & Characteristic radius of gas ejection for the diffuse component. & U[2.0,\,8.0] & 3.5  \\
\bottomrule
\end{tabular}%
}
\end{table*}

\subsection{Baryonification}
\label{sec:bfc_review}
Baryonification provides an efficient framework for incorporating the effects of baryons into dark matter-only simulations \citep{Schneider_2015, Schneider_2019, schneider_2025}. Rather than directly simulating galaxy formation, baryonification modifies halo density profiles using physically motivated dark matter, stellar, and gas components whose properties can be constrained by observations.  
In particular, a Navarro–Frenk–White (NFW)-like profile is converted into a baryonified profile consisting of a hot ionized gas component $\rho_{\rm hga}$, a central galaxy component $\rho_{\rm cga}$, a collisionless matter component  $\rho_{\rm clm}$ (i.e., dark matter, halo stars, and satellites), and the contribution from neighboring halos $\rho_{\rm 2h}$ (commonly referred to as the two-halo term):
\begin{equation}
    \rho_{\rm nfw}(r)  \rightarrow \rho_{\rm bfc}(r) = \rho_{\rm hga}(r) + \rho_{\rm cga}(r) + \rho_{\rm clm}(r) + \rho_{\rm 2h}(r).
    \label{eq:density_total_BC7}
\end{equation}

The impact of baryonic feedback is primarily encoded through the hot gas density profile.
The original baryonification prescription adopted an observationally motivated, single component gas profile: 

\begin{equation}\label{eq:gas_density_BC7}
\rho_{\rm hga}(r) = \rho_{{\rm hga},0}
\left(1+\frac{r}{r_{\rm c}}\right)^{-\beta(M_{200})}
\left[1+\left(\frac{r}{r_{\rm ej}}\right)^{\gamma}\right]^{-\frac{\delta-\beta(M_{200})}{\gamma}}\,,
\end{equation}
where $\rho_{{\rm hga},0}$ is the normalization, set by the halo's hot gas fraction $f_{\rm hga}$ (described below) and $r_{\rm c}$ is the core radius, following $r_{\rm c} = \theta_{\rm c} R_{200}$, with $\theta_{\rm c} = 0.1$. The truncation beyond the ejection radius, $r_{\rm ej}$, is controlled by $\gamma$ and $\delta$, governing the sharpness of the transition and the outer slope respectively. $\beta(M_{200})$ is a mass-dependent function parameterised as: 

\begin{equation} \label{eq:beta_model_BC7}
    \beta(M_{200}) = \frac{3\,(M_{200}/M_{\rm c})^{\mu}}{1 + (M_{200}/M_{\rm c})^{\mu}}\,.
\end{equation}
At large masses, this function asymptotically reaches three, and it goes to zero for low values of $M_{200}$. The characteristic mass, $M_{\rm c}$, sets the transition between the diffuse and centrally-concentrated limits. The parameter $\mu$ controls how fast this transition occurs.
The parameter $\theta_{\rm ej}$ determines the radial extent of the gas through $r_{\rm ej}=\theta_{\rm ej}R_{200}$. Together, these parameters regulate both the amount and the extent of the ejected gas.

The central galaxy component $\rho_{\rm cga}$ is commonly treated as an exponentially truncated power law.
The collisionless matter component $\rho_{\rm clm}$ is assumed to be a NFW profile.
At a given halo mass, the fraction of the baryon budget sequestered into stars ($f_\text{star}$) and the central galaxy ($f_{\rm cga}$) are set by the power-law relations of \cite{Moster2013}.
The satellite fraction is defined as $f_{\rm sga}=f_\text{star}-f_{\rm cga}$.
The hot gas fraction is likewise defined as $f_{\rm hga} = f_\mathrm{bar}-f_\text{star}$, where $f_{\rm bar} \equiv \Omega_{\rm b}/\Omega_{\rm m}$.

\subsection{Two-Component Model} 
\label{sec:two_component}

We propose to decompose the hot-gas profile into two components: an inner hydrostatic component and an extended, diffuse outer component. 
While each component could be made arbitrarily flexible, our goal is to develop the simplest physically motivated extension of baryonification. 
For the inner component, we assume a near-hydrostatic profile, and for the outer component, we only vary the radial extent of the distribution.
By keeping the two constituent profiles inflexible, the shape of the combined gas profile is determined simply by how much gas resides in the inner versus outer components.
This two-component approach is not intended as a precise physical picture.
Instead, it seeks to recast the flexible, single component model commonly used in baryonification in terms of physically motivated quantities.  

To parametrise the amplitudes of the two components, we introduce the fraction $f_{\rm comp}$, which specifies the fraction of the hot gas budget in the hydrostatic component. Parametrising the model in this way ensures conservation of the total hot-gas fraction, avoiding the degeneracies that would arise if the two component amplitudes were varied independently. We define the component fraction as
 \begin{equation}\label{fcomp_equation}
f_{\text{comp}}= \frac{\left(M_{200}/{M_{\rm c}}\right)^{\mu}}{1 + \left( M_{200}/{M_{\rm c}}\right)^{\mu}} \,.
 \end{equation}
At low masses ($M_{200} \ll M_{\rm c}$), the diffuse component is dominant, and at high masses ($M_{200} \gg M_{\rm c}$), the hydrostatic component dominates.
To draw an analogy with the standard baryonification framework, we refer to the transition mass as $M_{\rm c}$, and parametrise the sharpness of the transition with the power-law index, $\mu$.
The gas fractions of the two components are then 
\begin{equation}
\begin{aligned}
f_{\mathrm{h}} &= f_{\mathrm{hga}}\, f_{\mathrm{comp}} \\
f_{\mathrm{d}} &= f_{\mathrm{hga}} - f_{\mathrm{h}} \, , 
\end{aligned}
\label{eq:gas_components}
\end{equation}
where $f_{\mathrm{h}}$ is the mass fraction of the hydrostatic inner component, and $f_{\mathrm{d}}$ is the mass fraction of the diffuse outer component, both relative to the total halo mass. 

\begin{figure}
    \centering
    \includegraphics[width=\columnwidth]{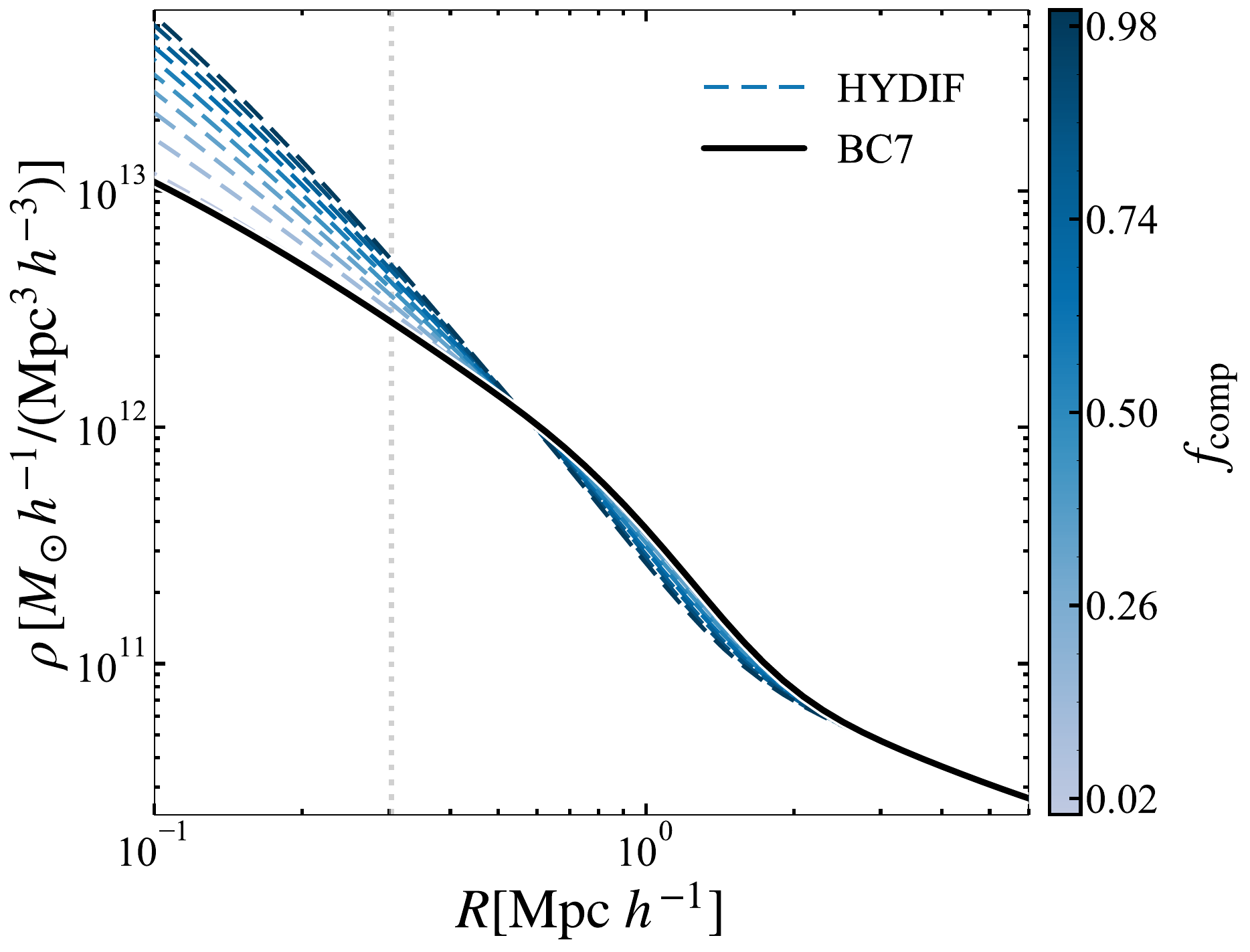}
    \caption{Gas density profiles for $M_{\rm 200}=2.19\times10^{13}\,M_\odot$, comparing the two-component model (\textsc{HyDif}, dashed) with the single-component model (BC7, solid). \textsc{HyDif} splits the gas into a dense, hydrostatic inner component and a diffuse ejected component, producing a steeper central profile and a sharper intermediate falloff than  BC7. The two converge in the outskirts where both models are dominated by the two-halo term. The vertical dotted line marks $R_{500}$. 
    % The parameters used for each models are fixed at their fiducial values: $\gamma = 2.5$, $\eta = \delta\eta = 0.2$, $\mu = 1.0$, $\theta_{\rm ej} = \theta_{\rm d} = 3.5$, $\delta_{\rm b} = \delta_{\rm d} = 8.0$, $\log M_{c, \ \rm d} = 13.32$, matching the single-component BC model's $\log M_{\rm c} = 13.32$, at $z = 0$. 
    For \textsc{HyDif}, we show the effect of varying the component fraction, as reflected in the colour bar.} 
    \label{fig:density_profiles}
\end{figure}

The combined hot gas density component is defined as 
\begin{equation}
    \rho_{\rm hga}(r)=\rho_{\rm h}(r)+\rho_{\rm d}(r)\, ,
\end{equation}
in terms of the hydrostatic, $\rho_{\rm h}(r)$, and the diffuse, $\rho_{\rm d}(r)$ components. Figure~\ref{fig:density_profiles} presents the two-component decomposition alongside the original single-component baryonification model. 

The shape of the hydrostatic component is
\begin{equation}
\rho_{\rm h}(r) \propto 
\left(1 + \frac{r}{r_{\rm c}}\right) ^{-\beta_{\rm h}}
\left[1+\left(\frac{r}{r_{\rm h}}\right)^{\gamma}\right]^{-\frac{\delta_{\rm h}-\beta_{\rm h}}{\gamma}},
\label{eq:rho_hydro}
\end{equation}
with a fixed shape according to: $\beta_{\rm h}=2.7$, $\gamma = 2.5$, $\delta_{\rm h} = 7$,  and $\theta_{\rm h}=3.5$ ($r_{\rm h} = \theta_{\rm h} R_{200}$). 
With the exception of $\beta_{\rm h}$, these are the fiducial values adopted by \cite{Giri_2021} after calibration against a suite of hydrodynamical simulations. 
The inner slope is set to $\beta_{\rm h} = 2.7$ rather than the asymptotic value, $\beta = 3$, of the standard parametrisation, because $\beta_{\rm h} = 3$ produces a gas profile concentrated enough to drive strong adiabatic contraction of the dark matter, raising the enclosed gas fraction above the cosmic baryon limit. A slightly shallower inner slope relaxes the contraction and restores a physically consistent baryon budget.

The shape of the diffuse outer component is
\begin{equation}
\rho_{\rm d}(r) \propto 
\left(1+\frac{r}{r_{\rm c}}\right)^{-\beta_{\rm d}(M_{200})}
\left[1+\left(\frac{r}{r_{\rm d}}\right)^{\gamma}\right]^{-\frac{\delta_{\rm d}-\beta_{\rm d}(M_{200})}{\gamma}}\,,
\label{eq:rho_diffuse}
\end{equation}
where we allow the radial extent of the component to vary through
$\theta_{\rm d}$ ($r_{\rm d} = \theta_{\rm d} R_{200}$).
The remaining parameters are held fixed: $\gamma=2.5$, $\delta_{\rm d}=8$, $M_{\rm d} = 10^{15} M_\odot$, and $\mu_{\rm d}=1$. 
The parameters $M_{\rm d}$ and $\mu_{\rm d}$ set the power-law $\beta_{\rm d}$ as
\begin{equation}
\beta_{\rm d}(M_{200}) = \frac{3\,(M_{200}/M_{\rm d})^{\mu_{\rm d}}}
{1 + (M_{200}/M_{\rm d})^{\mu_{\rm d}}}.
\label{eq:beta_diffuse}
\end{equation}
Because we seek a physically interpretable parametrisation, we keep the individual components inflexible and allow the shape of the combined profile to change through $f_\mathrm{comp}$ (the relative contribution of the hydrostatic term) and $\rm \theta_\mathrm{d}$ (the radial extent of the diffuse component).
The model therefore has only three free parameters:
$M_{\rm c}$, $\mu$, and $\theta_{\rm d}$. 

We adopt the central galaxy and collisionless matter components from \citet{Schneider_2019}. We implement an updated two-halo term profile based on \citet{schneider_2025}: to prevent overlapping halos, a halo-exclusion prescription suppresses the two-halo term inside the virial radius, where distinct nearby halos cannot physically contribute.

\begin{figure*}
    \centering
    \includegraphics[width=1 \textwidth]{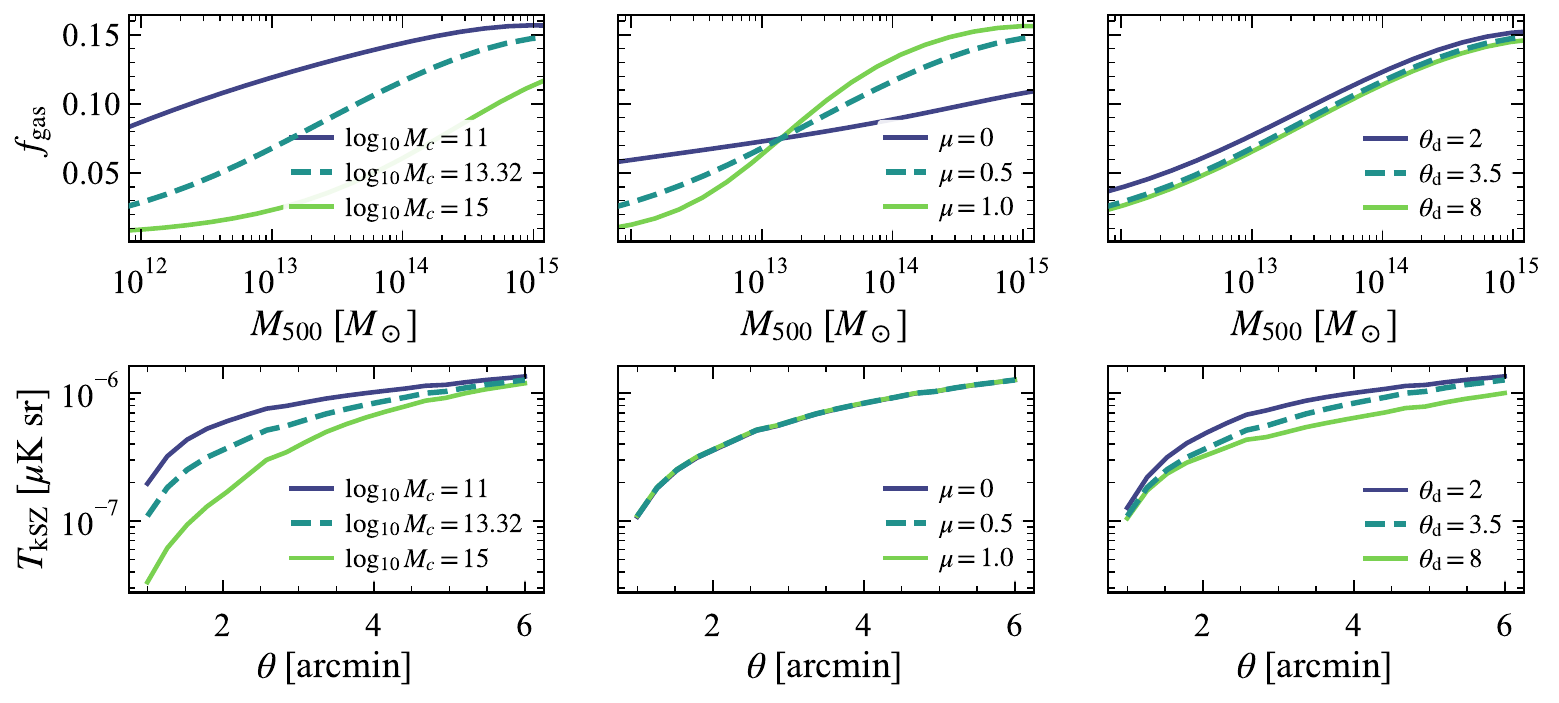}
    \caption{Sensitivity of the gas mass fraction (top) and kSZ (bottom) observables to the free parameters of the \textsc{HyDif} model. Each column varies a single parameter about its fiducial value (dashed curve) while holding the others fixed. The assumed values for each fixed parameter are the fiducial values listed in Table~\ref{tab:bc_update}.
    Here, the kSZ prediction is evaluated at a single halo mass ($M_\text{200} = 2.19 \times 10^{13} M_\odot$) close to $M_{\rm c}$, such that $\mu$ has little leverage on the kSZ profile; it is instead constrained by the mass dependence of $f_{\rm gas}$ (top row). }
    \label{fig:params_bc3}
\end{figure*}

\subsection{Modelling gas observables}
Here, we describe how predictions for the gas fraction and kSZ signal are computed from the halo density profiles. The response of the gas mass fraction and kSZ observations to the three \textsc{HyDif} free parameters is illustrated in Figure~\ref{fig:params_bc3}.

Halo gas mass fractions provide a measure of the amount of gas retained within halos (e.g. $R_{500}$). Since X-ray emission scales approximately as density squared, X-ray observations are primarily sensitive to the dense hydrostatic gas in the central regions. 
We compute the enclosed gas fraction by integrating the total gas density profile within $R_{500}$. 

Unlike X-ray emission, the kSZ effect depends linearly on the free-electron density, making it sensitive to the diffuse ionized gas extending beyond the hydrostatic core. For a known line-of-sight velocity, the kSZ signal therefore provides a direct probe of the electron distribution throughout the halo outskirts. The temperature fluctuation produced by the kSZ effect relative to the  present-day CMB temperature, $T_{\rm CMB}$, is
\begin{equation}\label{ksz_fluctuations}
\frac{\Delta T_{\text{kSZ}}(\bm{\theta})}{T_{\text{CMB}}}
= - \sigma_{\rm T} \int \frac{d\chi}{1+z} \,
n_{\rm e}(\bm{\theta}, z) \, e^{-\tau(z)} \,
\frac{v_{\mathrm{e},r}(\bm{\theta}, z)}{c}\, ,
\end{equation}
where $\sigma_{\rm T}$ represents the Thomson scattering cross-section, $\chi$ is the comoving radial distance, and ${v}_{\rm e}$ is the electron peculiar velocity projected along the line of sight. 
For group mass halos, the gas is optically thin ($\tau \ll 1$), therefore for one galaxy:
\begin{equation}\label{kSZ_with_cap}
\frac{\Delta T_{\text{kSZ}}(\bm{\theta})}{T_{\text{CMB}}}
\approx - \tau_{\text{gal}} \, \frac{v_{\mathrm{e},r}}{c} \, .
\end{equation}

The kSZ measurements considered in this work are stacks across many galaxies.
For a given galaxy, the temperature fluctuation profile is measured with compensated aperture photometry (CAP) filters, which remove large-scale fluctuations.
The CAP-filtered temperature is defined as
\begin{equation}\label{temp_fluctuations_capfilter}
T(\theta_{\rm ap}) = \int d^2 \theta \, \Delta T_{\rm kSZ}(\theta) W_{\theta_{\rm ap}}(\theta),
\end{equation}
where \( W_{\theta_{\rm ap}} \) is the CAP filter:
\begin{equation}\label{capfilter}
W_{\theta_{\rm ap}}(\theta) = 
\begin{cases} 
1 & \text{if } \theta < \theta_{{ \rm ap}} \\
-1 & \text{if } \theta_{\rm ap} \leq \theta \leq \sqrt{2} \theta_{\rm ap}, \\
0 & \text{otherwise}.
\end{cases}
\end{equation}
To predict the kSZ signal from a given gas density profile, we first calculate the electron density profile assuming a fully ionized medium and a hydrogen mass fraction of $0.76$.
We then obtain the optical depth by line-of-sight integration and apply both the instrumental beam and the CAP filter.

\section{Simulation study}\label{sec:sims}
The baryonification framework has previously been validated against the gas, dark matter, and stellar mass distributions of a range of cosmological hydrodynamical simulations \citep{Schneider_2019, schneider_2025}. We explore an extension to this framework, by replacing the single-component halo gas density profile with a two-component profile, representing inner hydrostatic and outer diffuse gas components;
we refer to this extension as \textsc{HyDif}.
To retain physical interpretability, we fix the shape of the inner hydrostatic component, and only allow limited flexibility in the outer component, such that the shape of the combined gas profile is set by the relative contribution of the hydrostatic and diffuse components.  We first examine whether this approach retains the ability to reproduce the diversity of gas density distributions found in simulations.  

Varying the three free parameters ($\log_{10} M_{\rm c}$, $\mu$, $\theta_{\rm d}$), we fit  \textsc{HyDif} to the hot gas halo density profiles extracted from five cosmological simulations: FLAMINGO \citep{Schaye_2023}, BAHAMAS \citep{McCarthy_2016}, SIMBA \citep{Dav__2019}, FABLE \citep{Henden_2018}, and XFABLE \citep{Bigwood_2025}.  These simulations differ in their subgrid implementations of black hole accretion and AGN feedback, resulting in diverse predictions for the gas density distributions in and around galaxy groups and clusters \citep[e.g.,][]{bigwood_ksz_2025}.  The simulated profiles are computed at $z=0$ from halos at $13.25<\log_{10}(M_{500} /M_{\odot})<13.35$.

\begin{figure*}
    %\centering
    \includegraphics[width=1\textwidth]{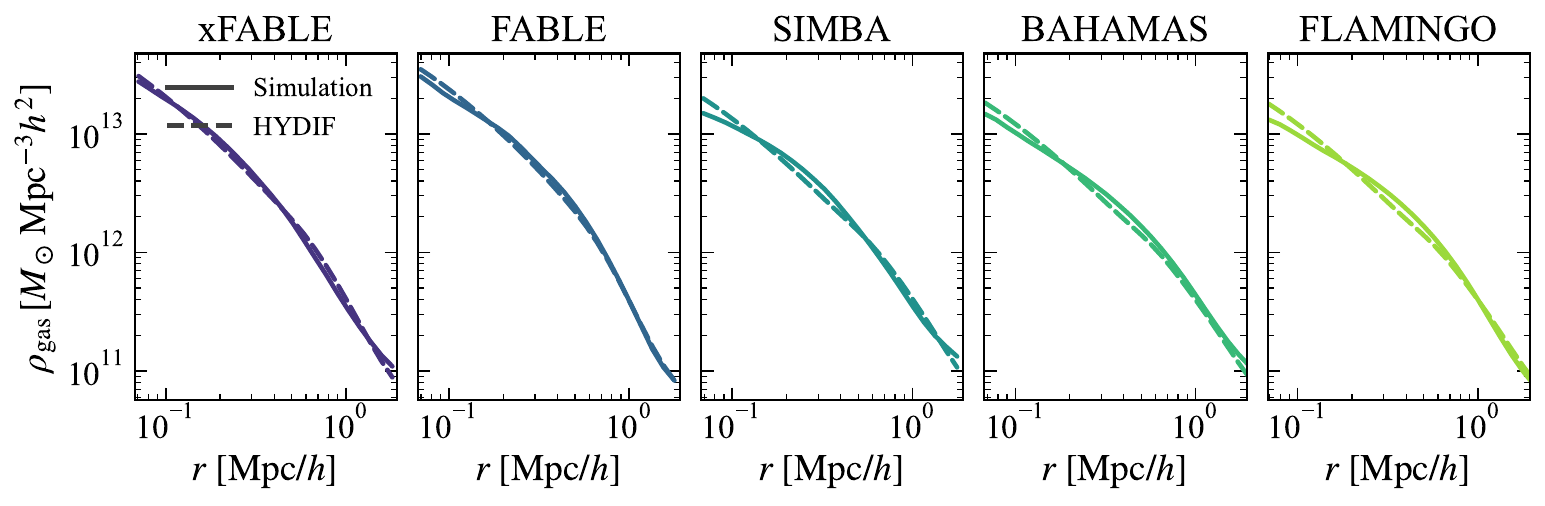}
    \caption{Hot gas profiles from hydrodynamical simulations (solid), alongside the best-fitting \textsc{HyDif} profiles (dashed). We consider \texttt{FLAMINGO} \citep{Schaye_2023}, \texttt{BAHAMAS} \citep{McCarthy_2016}, \texttt{SIMBA} \citep{Dav__2019}, \texttt{FABLE} \citep{Henden_2018}, and \texttt{XFABLE} \citep{Bigwood_2025}, which span a range of baryonic feedback implementations.}
    \label{fig:simulation_figure}
\end{figure*}

The best-fit \textsc{HyDif} profiles are shown in Figure~\ref{fig:simulation_figure}.  The two-component profile broadly reproduces the simulations' gas distributions, despite the different feedback implementations.  Further validation of the framework will require quantifying this performance across a wide range of halo masses and redshifts, which we leave to future work.

\begin{figure*}
    \centering
    \begin{minipage}[t]{0.48\textwidth}
        \centering
        \includegraphics[width=\linewidth]{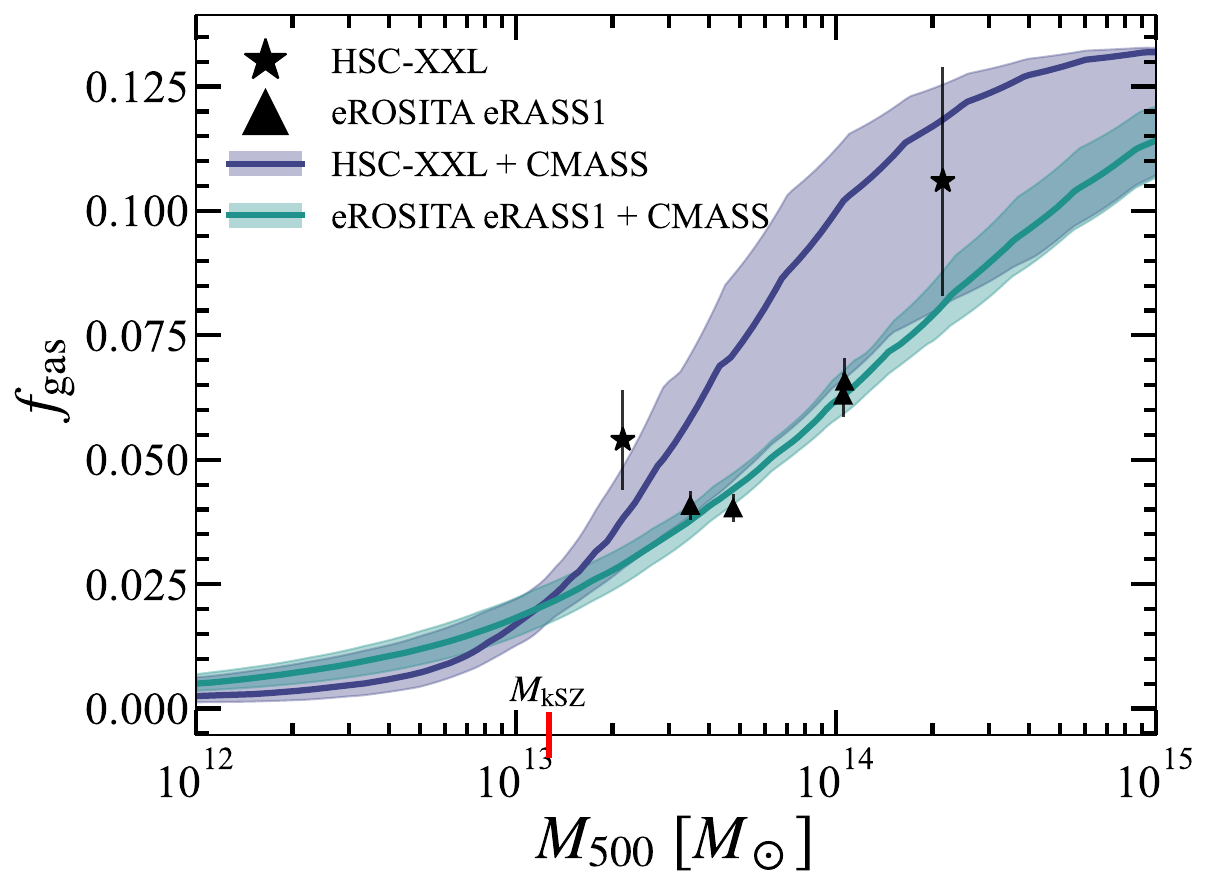}
    \end{minipage}
    \hfill
    \begin{minipage}[t]{0.48\textwidth}
        \centering
        \includegraphics[width=\linewidth]{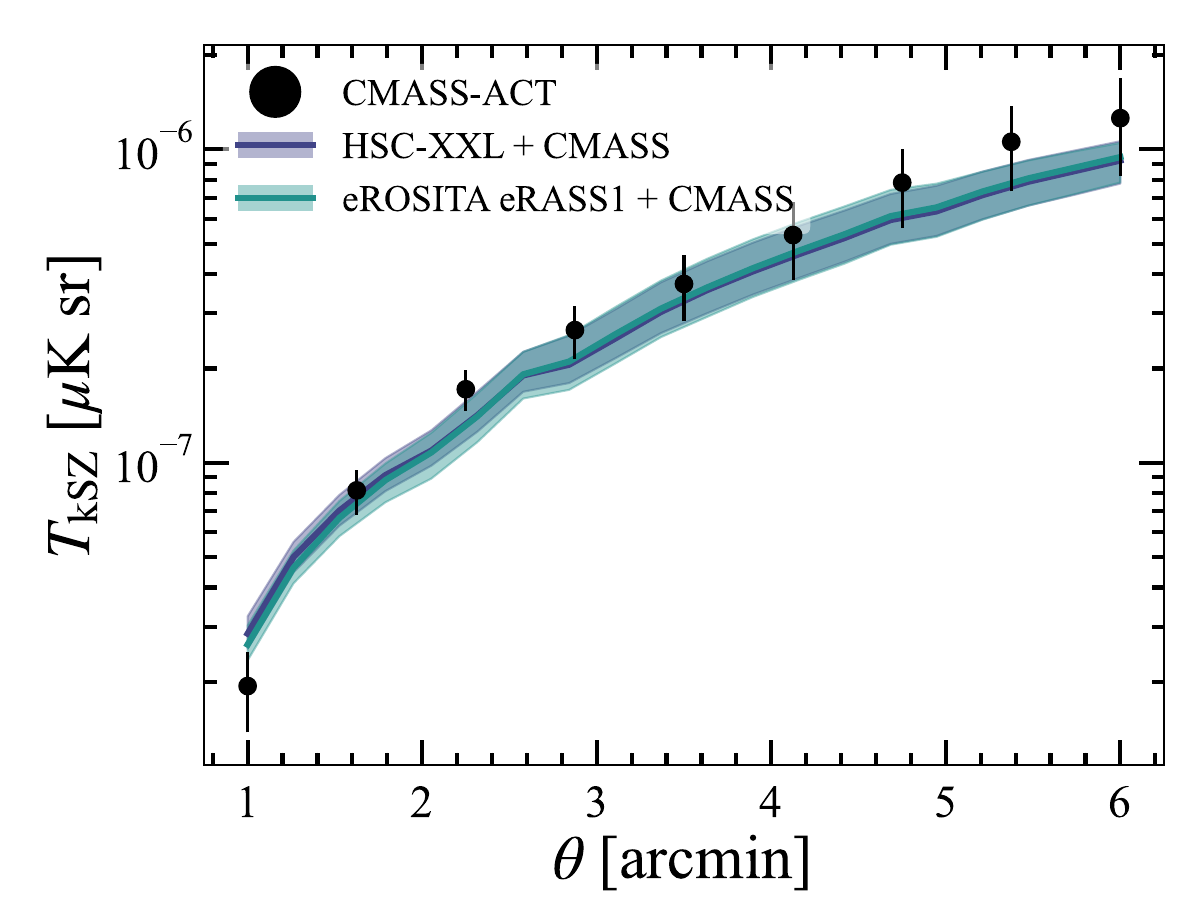}
    \end{minipage}
   \caption{Observational study:  Joint fits to  the CMASS kSZ measurements \citep{Schaan_2021} with two X-ray gas-fraction samples, HSC--XXL+CMASS (purple) \citep{Akino_2022} and eROSITA+CMASS (teal) \citep{siegel2025_flamingo}. \emph{Left:} The gas mass fractions, with the measurements shown as stars (HSC-XXL) and triangles (eROSITA). \emph{Right:} the kSZ signal, with the CMASS-ACT data shown in black. Solid lines show the posterior median and shaded bands the 68\% credible region. The two fits demonstrate the potential of the proposed two-component gas profile to retain flexibility.} \label{fig:fgas_ksz_fits}
\end{figure*} 

\section{Observational study}\label{sec:data}

In this work, we illustrate how decomposing halo gas into hydrostatic and diffuse components provides a simple, physically motivated framework for connecting complementary observations. 
In this section we demonstrate this idea through proof-of-concept modelling of X-ray and kSZ measurements.

\subsection{Data}

\subsubsection{kSZ measurements}
We consider the kSZ measurements of \citet{Schaan_2021}, which stacked Atacama Cosmology Telescope (ACT) CMB temperature maps at the locations of CMASS galaxies from the Baryon Oscillation Spectroscopic Survey (BOSS), using reconstructed peculiar velocities. The CMASS sample has a mean redshift of $z\simeq0.55$ and probes group-scale halos; here, we assume an average halo mass of $\log_{10}[M_{200}/(h^{-1}\,{\rm M_\odot})] = 13.34\pm 0.04$ \citep{mccarthy2025, siegel2025_flamingo}. This value is derived by fitting simulated galaxy samples to GGL measurements of the same CMASS data \citep{Amon2023}. 
We consider the higher sensitivity and higher resolution  measurement from the ACT $150$~GHz channel; the beam is approximately Gaussian with FWHM of $1.3$~arcmin.

\subsubsection{X-ray Gas Fractions}

X-ray gas fractions probe the amount of hot gas retained within halos and are therefore widely used to calibrate models of baryonic feedback. Recent studies have noted significant differences between the mean X-ray gas mass fractions reported from different surveys and analysis methods.
Gas mass fractions inferred from the eROSITA All-Sky Survey, including the publicly available eRASS1 catalogue of X-ray detected groups and clusters \citep{Bublul_2024,Dev2024,siegel2025_flamingo} and stacking measurements on optically selected groups \citep{Popesso_2024}, are systematically lower than previously found \citep[e.g.,][]{Eckert_2016,Akino_2022}. 
This discrepancy has important implications for baryonification models, leading to substantially different inferred feedback strengths and matter power spectrum suppression \citep{kovac_2025, siegel_2025}. 
Selection effects have been proposed as a potential explanation for the apparent discrepancy, however, the situation is not yet fully understood \citep[e.g.,][]{Seppi2022,Popesso2024a,Dev2024,Marini2024,Clerc2024,Seppi2025, Bucko2026}.
Given the uncertain landscape, we conservatively consider two sets of gas mass fraction estimates. 

XXL is a sample of over $300$ galaxy groups observed with \textit{XMM-Newton} at redshift $0<z<1$ with halo masses of $M_{500}\sim10^{13}$--$10^{15}\,\mathrm{M_\odot}$ \citep{Pierre_2016, Adami_2018}. 
A subset of the halos were characterized with weak lensing from HSC \citep{Umetsu_2020}.
We consider the gas fraction measurements of \citet{Akino_2022}, which reanalyzed the lensing-characterized XXL clusters and modelled the X-ray selection function.

We also consider the gas fraction measurements from the first eROSITA All-Sky Survey release (eRASS1)  \citep{Bublul_2024, Kluge_2024, Popesso_2024}. 
We adopt the four halo-mass and redshift bins analysed by \citet{siegel2025_flamingo}, which re-derived the mean halo mass of each bin with new galaxy--galaxy lensing measurements.
These GGL-calibrated bins are consistent with the stacked X-ray gas mass fractions of optically selected halos \citep{Popesso_2024}, as well as the recent analysis of the eROSITA Final Equatorial-Depth Survey sample \citep{Bucko2026}.

\subsection{Modelling Results}

We perform a joint Bayesian fit to the X-ray and kSZ data using \texttt{emcee} \citep{Foreman-Mackey_2013}. We vary the three profile parameters: $\log_{10}M_{\rm c}$ and $\mu$, which set the hydrostatic component fraction (Eq.~\ref{fcomp_equation}), and $\theta_{\rm d}$, which sets the radial extent of the diffuse, outer component. 

The fits are presented alongside the data in Figure \ref{fig:fgas_ksz_fits}. 
The full posterior distributions are shown in Appendix \ref{app:erosita_akino_contours}. We note that the goal of this study is to demonstrate that despite using only three free parameters, the two component approach retains considerable flexibility, such that it is able to reasonably describe different data, while offering added interpretability compared to the original baryonification parametrisation. 
The CMASS kSZ signal is matched across the full angular range for both joint X-ray fits;
however, we stress that the signal is dominated by the innermost radial bins due to the covariance of the measurement \citep{Schaan_2021}, and note that modelling the larger radii is more sensitive to systematics \citep{siegel_2025}.
For the joint kSZ and eROSITA X-ray fit, the model describes the gas mass fractions remarkably well.
The joint kSZ and HSC--XXL X-ray fit marginally struggles to reproduce the lower-mass X-ray point, an issue previously noted by several studies \citep{bigwood_2024,kovac_2025,siegel_2025}.
We note that the discrepancy between the pre-eROSITA X-ray gas mass fractions, such as HSC--XXL, and the new kSZ effect measurements is greatest for the SDSS LOWZ/ACT kSZ profile \citep{Schaan_2021}, which probes the same mass, redshift, and radial scale as the  HSC--XXL sample \citep{siegel_2025}.
In this work, we limit our focus to the description and demonstration of the two-component model, leaving discussion of the consistency of X-ray and kSZ measurements to future work.

\section{Discussion}\label{sec:discussion}

\textbf{A physically interpretable framework:} 
Existing baryonification approaches typically represent the gas distribution with a single effective component whose properties are adjusted to reproduce the net impact of baryonic feedback. Here we present a natural decomposition of the gas distribution into two components: a dense, approximately hydrostatic component, and a flexible, diffuse outer component. Although the implementation presented here is intentionally simple, this decomposition provides parameters with more direct physical interpretation and reflects the different physical environments probed by current observations.

The rapidly expanding observational landscape is ideal for constraining such a two-component model: X-ray gas fractions primarily constrain the dense hydrostatic gas, while the kSZ effect probes diffuse free electrons beyond the central halo. The framework therefore provides a natural language for combining multi-wavelength observations while retaining a clear connection to the underlying gas physics.

The implementation adopted here demonstrates the promise of this conceptual picture for reproducing both simulations and current observations. Despite varying only three physically motivated feedback parameters, \textsc{HyDif} retains much of the flexibility of the original baryonification framework, reproducing the diversity of gas density profiles in hydrodynamical simulations and providing a joint description of observed X-ray gas fractions and kSZ measurements. We emphasize, however, that this agreement should be viewed as a proof of concept rather than the endpoint of baryonic feedback modelling. 

\textbf{Towards next-generation baryonic feedback models:}
The \textsc{HyDif} implementation presented here demonstrates the potential of a hydrostatic/diffuse decomposition as a physically motivated foundation for future models. Increasingly flexible descriptions of  distinct gas reservoirs can be incorporated within a full forward model of the matter distribution.

In this work, we focus exclusively on the hot gas component of the baryonification model. The remaining components, stars, the central galaxy, and dark matter, are inherited from baryonification and are held fixed throughout. While this isolates the impact of the hydrostatic-diffuse decomposition, these components remain coupled through the global baryon budget, and variations in the stellar distribution, for example, will indirectly modify the hot-gas fraction. A complete implementation of  \textsc{HyDif} will therefore require simultaneously varying all baryonic components, as in previous baryonification studies, to construct self-consistent predictions for the matter distribution and weak-lensing observables.

\textbf{Future observational constraints:}
The two-component model is well suited to the next generation of multi-wavelength observations of halo gas. In this work, we considered X-ray gas fractions and kSZ measurements, which will substantially improve with future observations from \textit{eROSITA} \citep{Bublul_2024} and the Simons Observatory \citep{Ade_2019}. Fast radio bursts will provide a complementary probe of diffuse electrons extending into halo outskirts, while weak gravitational lensing and improved measurements of stellar mass fractions will continue to constrain the total matter and stellar components of the baryon budget. A physically interpretable framework naturally accommodates these complementary datasets, allowing each observable to constrain the gas component to which it is most sensitive while maintaining a consistent description of the halo baryon distribution.

\section{Conclusion}\label{sec:conclude}
In summary, we have introduced a physically motivated model that decomposes halo gas into a central hydrostatic core and an extended diffuse component, providing a simple and interpretable description of how baryonic feedback redistributes gas within dark matter halos. We demonstrated that this conceptual framework is sufficiently flexible to reproduce the hot-gas distributions of modern hydrodynamical simulations and to jointly describe current X-ray gas-fraction and kSZ observations using only three physically motivated feedback parameters.

The implementation presented here is intended as a proof of concept. The hydrostatic/diffuse decomposition provides a foundation upon which future baryonification models can be built, combining physically interpretable gas components with increasingly flexible parameterizations and full predictions for the matter distribution. As multi-wavelength observations dramatically improve over the next decade, this framework can provide a natural way to jointly constrain the distribution of baryons across dark matter halos and to incorporate these constraints into precision cosmological analyses.

\section*{Acknowledgements}
We thank Michael Kovac for helpful feedback and insights into the baryonification model.  

\section*{Data Availability}
The FLAMINGO, BAHAMAS, SIMBA, FABLE, and XFABLE simulations are all publicly available.

\bibliographystyle{mnras}
\bibliography{main} 

\appendix

\section{Posterior Distributions}
\label{app:erosita_akino_contours}

\begin{figure}
    \centering
    \includegraphics[width=\linewidth]{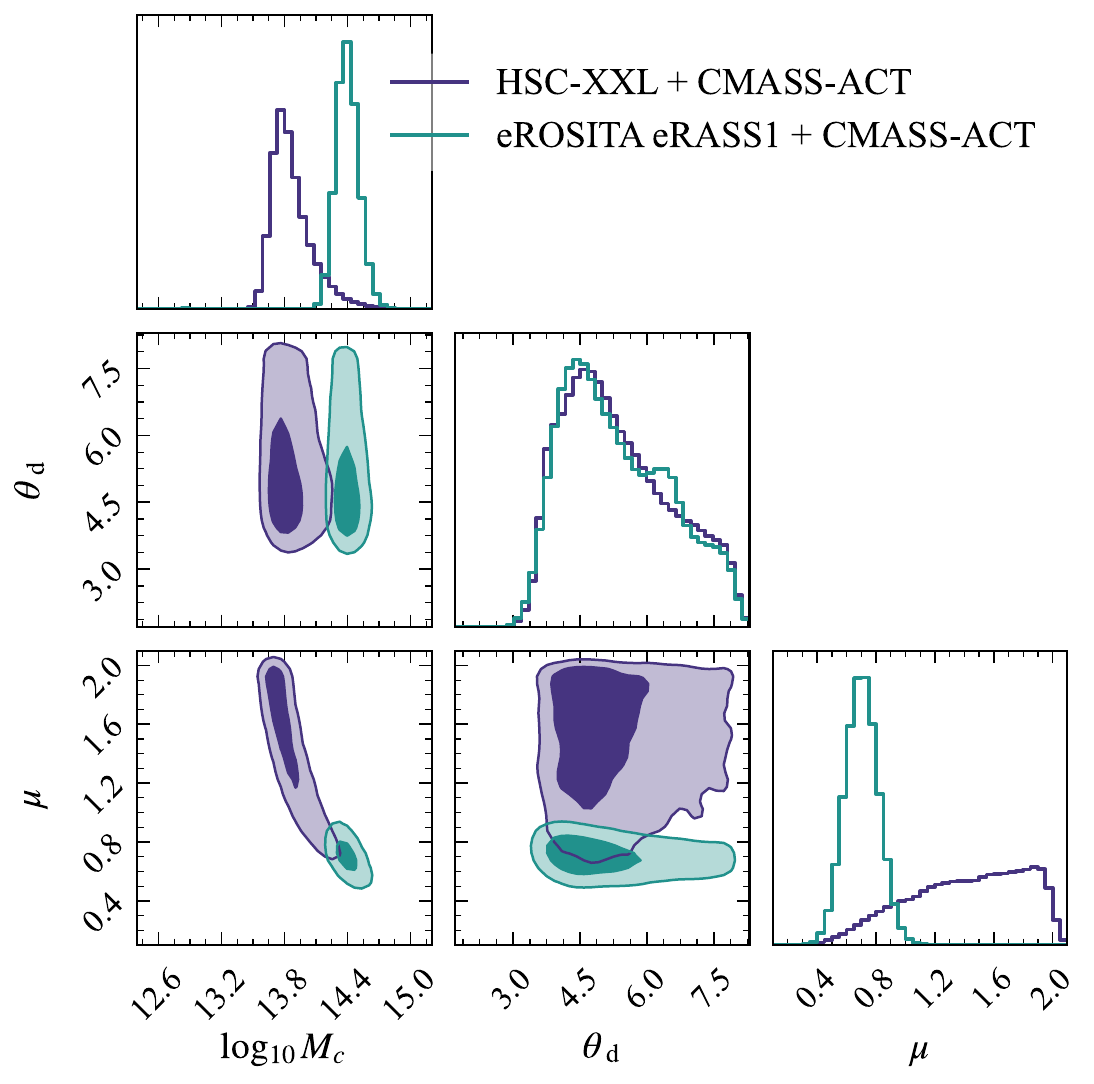}
    \caption{Posterior distributions for the three free \textsc{HyDif} parameters from the joint fit, comparing two X-ray gas-fraction datasets paired with the CMASS kSZ measurements: HSC-XXL+CMASS (purple) and eROSITA+CMASS (teal). Contours show the 68\% and 95\% credible regions. The kSZ-controlled parameter $\theta_{\rm d}$ is consistent between the two fits, as expected given the shared CMASS data, while the choice of X-ray sample shifts the hydrostatic-component parameters $\log_{10}M_{\rm c}$ and $\mu$, which trade off along a degeneracy direction. }
    \label{fig:erosita_akino_contours}
\end{figure}

Figure \ref{fig:erosita_akino_contours} shows the posteriors for the three \textsc{HyDif}
parameters under the joint HSC-XXL\,+\,CMASS fit and the eROSITA\,+\,CMASS fit. For HSC-XXL\,+\,CMASS, we find $\log_{10} M_{\rm c} = 13.81^{+0.22}_{-0.11}$, $\theta_{\rm d} = 5.15^{+1.59}_{-1.02}$, and $\mu = 1.43^{+0.40}_{-0.50}$.
The eROSITA\,+\,CMASS fit  returns $\log_{10} M_{\rm c} = 14.38^{+0.09}_{-0.08}$, $\theta_{\rm d} = 5.14^{+1.49}_{-1.0}$, and $\mu = 0.69^{+0.11}_{-0.11}$. The eROSITA\,+\,CMASS fit prefers a higher characteristic mass and correspondingly lower $\mu$, tracing
the same $\log_{10} M_{\rm c}$--$\mu$ degeneracy, while $\theta_{\rm d}$ is consistent between the two datasets.

\bsp	 
\label{lastpage}
\end{document}